\documentclass[epj,referee]{svjour}
\usepackage[dvips]{graphicx}

\def\EQ{\begin{equation}}
\def\EN{\end{equation}}
\def\EQA{\begin{eqnarray}}
\def\ENA{\end{eqnarray}}
\begin{document}

\title{A stochastic model of torques in von Karman swirling flow}
\author{N. Leprovost \and L. Mari\'e \and B. Dubrulle}
\institute{ Groupe Instabilit\'e et Turbulence,
\\CEA/DSM/DRECAM/SPEC and CNRS. URA 2464 F-91191 Gif sur Yvette Cedex, France
\\ \email{bdubru@drecam.saclay.cea.fr}}

\date{Received: date / Revised version: date}

\abstract{A stochastic  model is derived to predict the
turbulent torque produced by a swirling flow. It is a simple Langevin
process, with
a colored noise. Using the unified colored noise approximation, we
derive analytically the PDF of the fluctuations of injected
power in two forcing regimes: constant angular velocity or constant
applied torque. In the limit of small velocity fluctuations and
vanishing inertia, we predict that the injected power fluctuates twice
less in the case of constant torque than in the case of constant
angular velocity forcing.
The model is further tested against
experimental data in a
von Karman device filled with water. It is shown to allow for a
parameter-free prediction
of the PDF of power fluctuations in the case where the forcing is
made at constant torque. A physical interpretation of our model is
finally given, using a quasi-linear model of turbulence.}


\maketitle

\section{Introduction}
\label{intro}
\subsection{Historical background}
A classical topic in turbulence research is the computation of global
transport properties connected with the macroscopic result of
turbulent motions at microscopic scales. These motions are
characterized by very rapid characteristic time scales, and can be
considered, from a macroscopic point of view, as fluctuations. The "conventional approach" consists in modelling the mean value of non linear functions of these fluctuations in terms of averaged quantities, the most famous example being the turbulent viscosity \cite{Lesieur94}. However, this will only give evolution equations for averaged quantities and nothing can be predict on the shape of the fluctuations (for example their probability distribution function).  An alternative approach is to consider these fluctuations as noises and most of problems dealing with turbulent transport could then be solved if
one were able to prescribe the statistics of this noise, as a
function of some global properties of the flow. A priori, this can be
done in two ways. Firstly, by assuming the probability density function of the noise to be known. This approach has been pioneered by \cite{Hopf52}, starting from Navier-Stokes equations,
but its solution has encountered considerable technical difficulties
\cite{Monin77}. In some instances (e.g. for velocity
increments) where the noise obeys a Markov property, it is however
possible to derive an approximate Fokker-Planck equation by fitting
of the turbulent data 
\cite{Friedrich97a,Friedrich97b,Naert98,Marcq01}. Another way to 
prescribe the statistics of the noise is through
a stochastic equation, taking for example the Langevin equation.  In
some sense, this approach has been pioneered by \cite{Obukhov59}, who
assumes a Gaussian white noise statistics for the acceleration.
Refinements of this model have later been proposed by 
\cite{Castaing90,Delour01,Friedrich03,Beck03} to account for 
intermittency
of small scale velocity increments.\

\subsection{The Langevin approach}
From a practical (numerical) point of view, Langevin approaches are
often easier to implement, since they only involve integration of
ordinary differential equation, in contrast with Fokker-Planck methods
which involve partial differential equation in a high dimension space
(the phase space). From the theoretical point of view, the link
between the two approaches is not straightforward: it can be shown
that different Langevin models can in fact lead to the same
Fokker-Planck equation, so that the correspondence between the noise
property and the fluid equations of motions is not always obvious. In
that respect, it would be interesting to develop some sort of
systematic procedure to derive a Langevin equation for the noise
starting from the Navier-Stokes equations.  Recently, \cite{Laval99,Carlier01,Laval01} provided
evidence that the small scales of a turbulent flow are mostly slaved
to the large scale, and follow a
quasi-linear dynamic
\cite{Dubrulle97,Nazarenko99,Laval00,Dubrulle01}. This dynamics is
described by the rapid Distortion Theory, see e.g.
\cite{Townsend76,Keffer78,Maxey81,Kevlahan93,Kevlahan97}. This led us to
propose a new turbulent model for small scale turbulence, in which
the velocity is given as a solution of a linear stochastic equation
of Langevin type \cite{Laval03}. A preliminary validation of the
model was done by comparison with direct numerical simulation of
isotropic 3D turbulence. This kind of turbulence is however seldom
realized in real life applications. Therefore, a validation of this
type of model in non-isotropic, non-homogeneous situations would be
most welcome.\

\subsection{A model experiment}
\begin{figure}[ht]
\begin{center}
\includegraphics[scale=0.6]{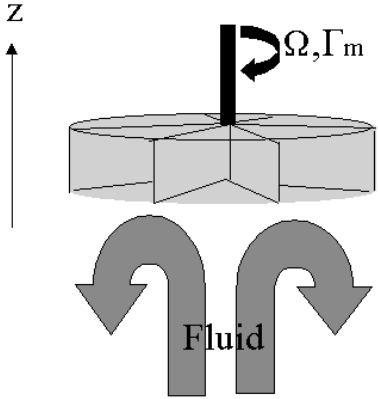}
\caption{\label{VKdevice} Von Karman experimental setup, $\Omega$ and
$\Gamma_m$ are respectively the angular velocity of the disk and the
torque supplied by the engines.}
\end{center}
\end{figure}
A good prototype of this type of flow is the so-called von Karman
flow, the flow between two coaxial rotating disks (cf figure \ref{VKdevice}). This simple device allows
both for  turbulence with a very large Reynolds number and easy
access to global transport properties via torque measurements
focusing either on averaged quantities
\cite{Labbe96b,Cadot97,Aumaitre00} or probability distributions
\cite{Titon03a}. The statistical analysis revealed a rich and
complex connection between the energy injection and dissipation,
reflecting the non-trivial coupling between the macroscopic scale and
the underlying microscopic turbulent noise.
For example,
Titon and Cadot \cite{Titon03a} studied power injection statistics, in
the regime obtained  when the disks are counter-rotating at same
angular velocity. In this case, the stationary state is made of two
cells with opposite azimuthal velocity.
The measurements of Titon and Cadot cover a range of Reynolds number
between $2\times 10^4$ to $5\times 10^5$, in two regimes: in the
first one, the angular velocity of the stirrers is constant ($\Omega$-mode); in the
second one, the mechanical torque is kept constant in time ($\Gamma$-mode). 
For each mode, the shape of the injected power statistics is found independent
of the Reynolds number. It is approximately Gaussian, with a slight
asymmetry. While the rate of fluctuations of the injected power is
independent of the Reynolds number within each mode, it is found to
depend strongly on the type of mode: it is twice larger in the first
regime than in the second one.\

In stationary regime, one expects the energy injection to be equal on
average to the energy dissipations. Yet, the two processes clearly
differ: the nearly Gaussian character of the PDF's of energy
injection fluctuations contrasts with the very non-Gaussian
(log-normal) behavior observed for energy dissipation. Also, the
strong dependence of the statistics on the forcing mechanism goes
against the universality assumption usually applied on energy
dissipation in classical theories of turbulence. These interesting
differences are far from being completely understood, from a
theoretical point of view. In a recent work, Aumaitre et al
\cite{Aumaitre01} showed that the statistics of the injected power
obey a "fluctuation theorem", enabling to connect the probabilities
of positive and negative production rate during a given time interval
(this characterizes the asymmetry of the curve). However, a rigorous
proof of the theorem only applies to time-reversible systems, at
variance with ordinary turbulence. Aumaitre et al \cite{Aumaitre01}
therefore also mention that their result could be just a consequence
of the theory of large deviations \cite{Ellis85,Oono89}.\

\subsection{Aim of the paper}
These two examples illustrate clearly the complexity of the global
transport properties occurring in the von Karman device. The
questions we address in the present paper are: i) can we capture the
main features of the transport through a simple Langevin model ? ii)
can we make a link between this Langevin model and the Navier-Stokes
equations through the quasi-linear model of turbulence of
\cite{Dubrulle97,Nazarenko99,Laval00,Dubrulle01,Laval03} ?\

We answer to these questions in two separate Sections, one devoted to
the finding and analyzing of the Langevin model, and one devoted to
its possible justification through the turbulent model. To further
test the basic hypothesis of the model, and to validate it
thoroughly, we used confrontation with experimental data collected
specifically for this purpose in the von Karman experimental device
of Saclay, described in \cite{Marie03b}.

\section{$\Omega$-mode and the Langevin model}
\subsection{Momentum equation}
\label{Omegamode}
To derive the simplest Langevin model compatible with the data, we may
follow Titon and Cadot \cite{Titon03a}, and write the momentum
balance equation for
one stirrer (including blades and water trapped in it), as :
\begin{equation}
\label{evolution}
I\frac{d\Omega}{dt} = \Gamma_m(t) - \Gamma_f(t)
\end{equation}
where $I$ is the inertia of the disks (including the blades and the
water trapped in it), $\Omega$ is the rotation velocity of the disk,
$\Gamma_m$ is
the angular momentum supplied by the motors (the propeller) and
$\Gamma_f$ is the torque due to the fluid acting onto the propeller.

\subsection{Derivation of the Langevin model}

\begin{figure}[ht]
\begin{center}
\includegraphics[scale=0.42,clip]{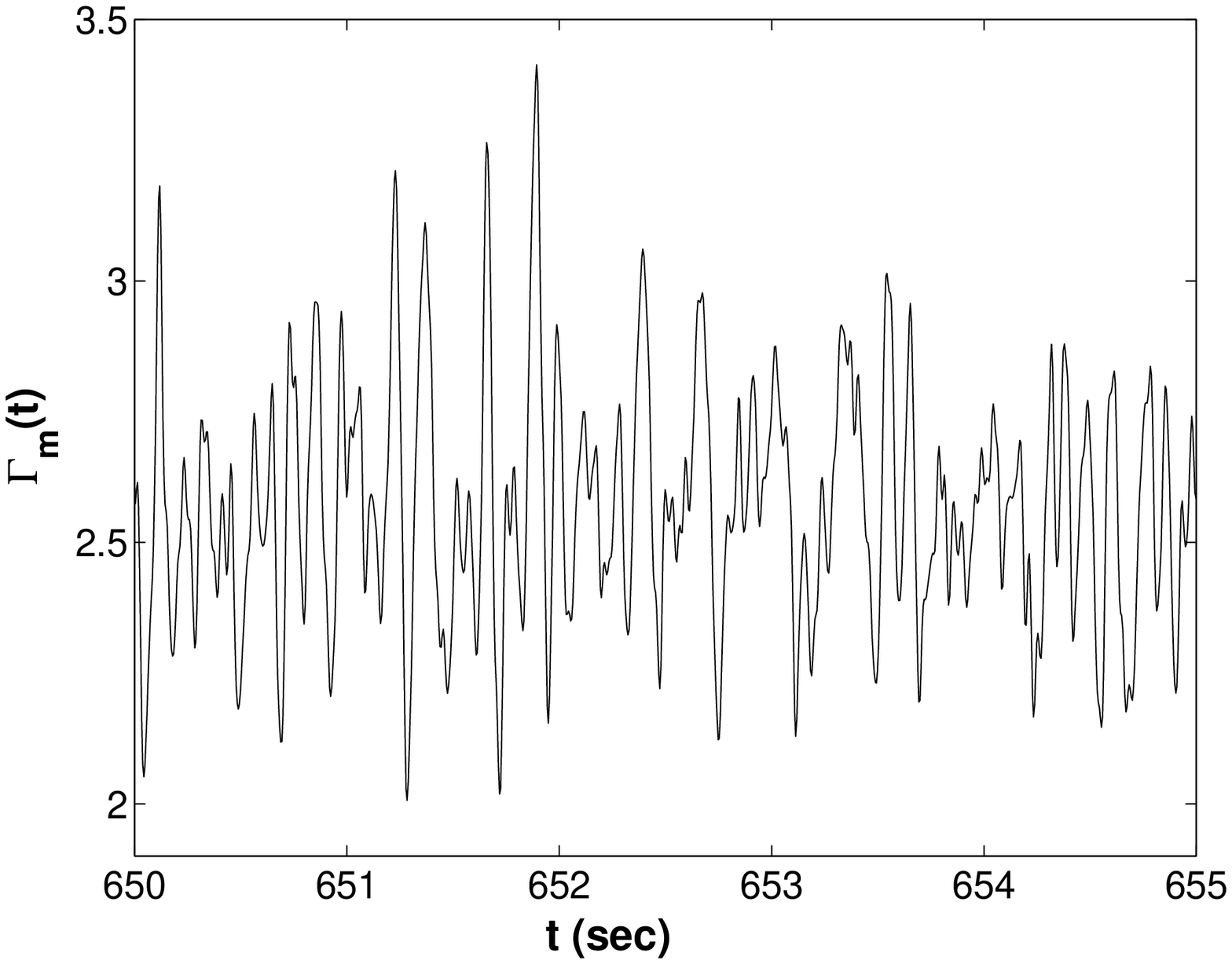}
\vspace{0.5cm}
\includegraphics[scale=0.42,clip]{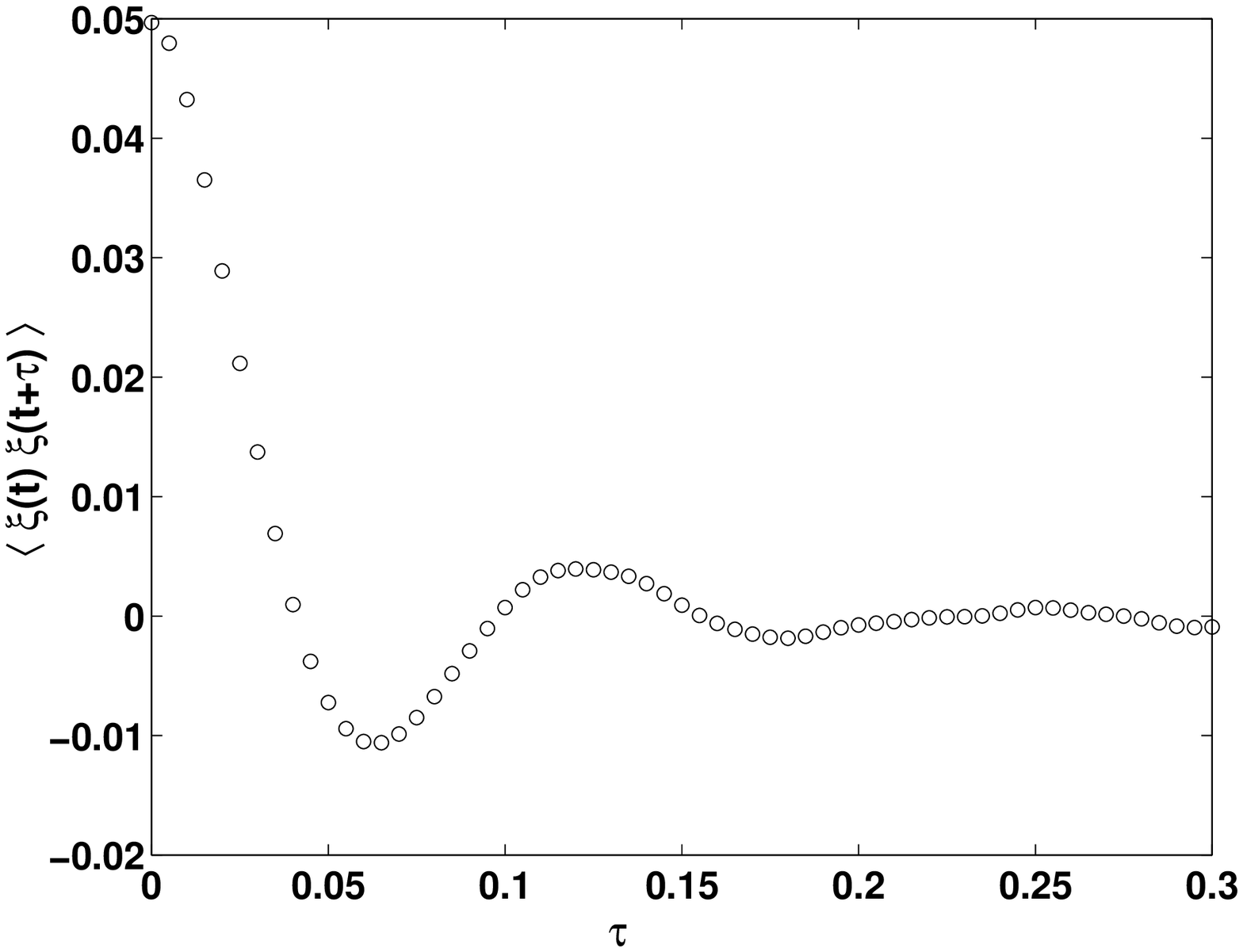}
\caption{\label{Correlation} Sketch of 5 seconds of torque signal from the Saclay experiment (top) and temporal correlation (bottom).}
\end{center}
\end{figure}

\begin{figure}[ht]
\begin{center}
\includegraphics[scale=0.42,clip]{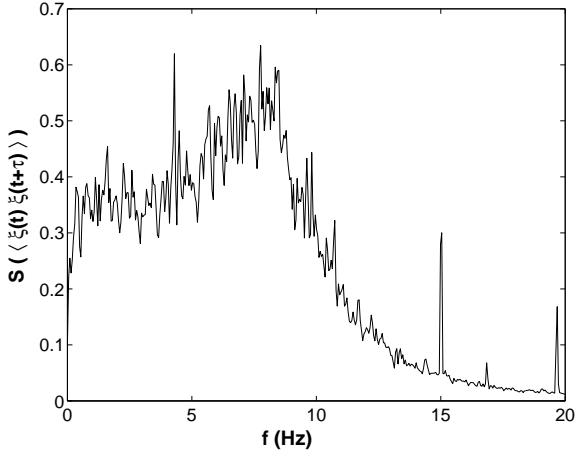}
\caption{\label{Spectre} Spectrum of the correlation function.}
\end{center}
\end{figure}

  From a theoretical point of view, $\Gamma_f(t)$ is the turbulent
contribution to be modeled as a noise. Its main properties can be
easily specified by working in the $\Omega$-mode, where
$\Gamma_m=-\Gamma_f$, and study the signal delivered by the motor. From experiments performed with a
von-Karman device working in water, regulated in $\Omega$-mode
($\Omega = 59.6 \;rad.s^{-1}$), one observe a roughly Gaussian distribution with a mean value
proportional to $\Omega^2$, with prefactor having the sign of
$\Omega$. We thus write $\Gamma_f = c \Omega \vert \Omega \vert -
\xi$ where $\xi$ is a Gaussian noise with zero mean, specified by its second order moment (the variance). To completely determine it, we extracted a temporal correlation for
the mechanical torque (figure \ref{Correlation} shows this
correlation versus time). One sees an oscillation at a frequency
around $8.9 \; Hz$ ($56.2 \;rad.s^{-1}$), superposed to a rapid
damping.
On the other hand, the Fourier transform of the signal displays a
rather wide ranging from $0$ to about $9 \; Hz$  (Figure \ref{Spectre}),
instead of a well defined narrow peak, which would be characteristic
of a meaningfull oscillation. Finally, we note that such oscillation is not
visible in similar measurements performed in air \cite{Labbe96b}. It 
is therefore difficult to decide whether this oscillation is a
real physical feature, or provoked by some experimental artefact.

If the oscillating behavior of the correlation function is considered, the simplest Langevin model is the OWN (oscillating white noise) model:
\EQ
\frac{d^2 \xi}{dt^2} = -2\gamma \frac{d\xi}{dt} - \omega_0^2 \xi +
\Gamma(t) \;,
\label{riri}
\EN
where $\langle \Gamma(t) \Gamma(t') \rangle = 2 D_0 \delta (t-t')$. This equation leads to a
stationary gaussian distribution for $\xi$ (thermal equilibrium) with
variance $\langle \xi^2 \rangle = D_0 / (2 \gamma \omega_0^2)$ and a
temporal correlation $C(\tau)=\langle \xi(t+\tau) \xi(t) \rangle$
which reads:
\EQ
C(\tau) = e^{-\gamma \tau} [A \cos(\sqrt{\omega_0^2 - \gamma^2}
\tau)+B \sin(\sqrt{\omega_0^2 - \gamma^2} \tau)]
\EN
where $A=\langle \xi^2 \rangle$ and $B=\gamma A / 
\sqrt{\omega_0^2-\gamma^2}$.

If, on the other hand, the oscillation is a spurious experimental
artefact, the only physical property to take into account is the
damping, which appears approximately exponential. The resulting
Langevin equation is the EWN (exponential white noise) model:
\EQA
\frac{d\xi}{dt} &=& -\frac{1}{\tau}\xi +\frac{\eta(t)}{\tau}\, ,  \\ \nonumber
\langle \eta(t) \eta(t') \rangle &=& 2 D \delta(t-t')
\label{equaxi}
\ENA
This leads to an exponentially decaying correlation function
\EQ
C(\tau) = \frac{D}{\tau} \; e^{-\frac{t}{\tau}}
\EN

\subsection{Calibration}
\begin{figure}[ht]
\begin{center}
\includegraphics[scale=0.42,clip]{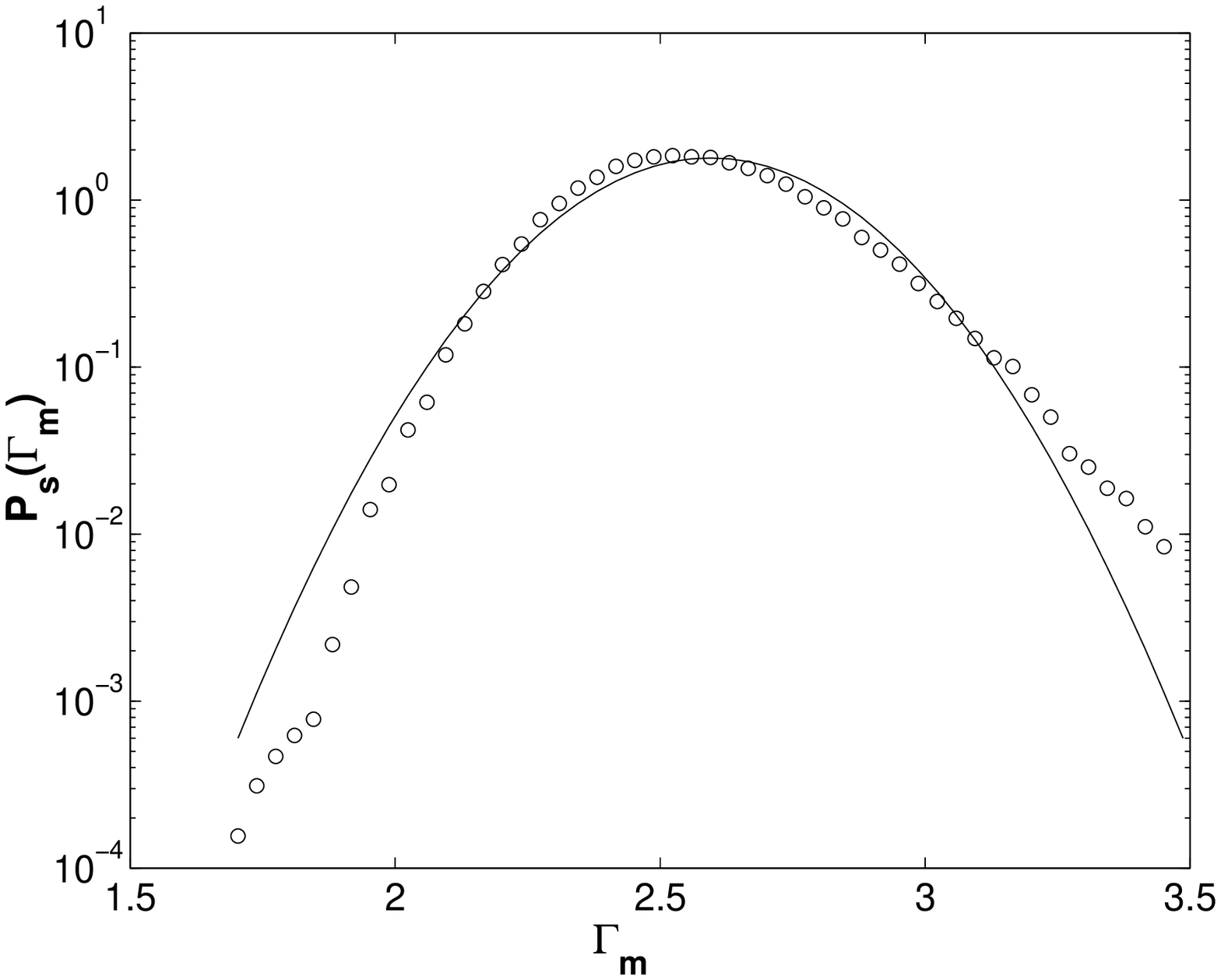}
\vspace{0.5cm}
\includegraphics[scale=0.42,clip]{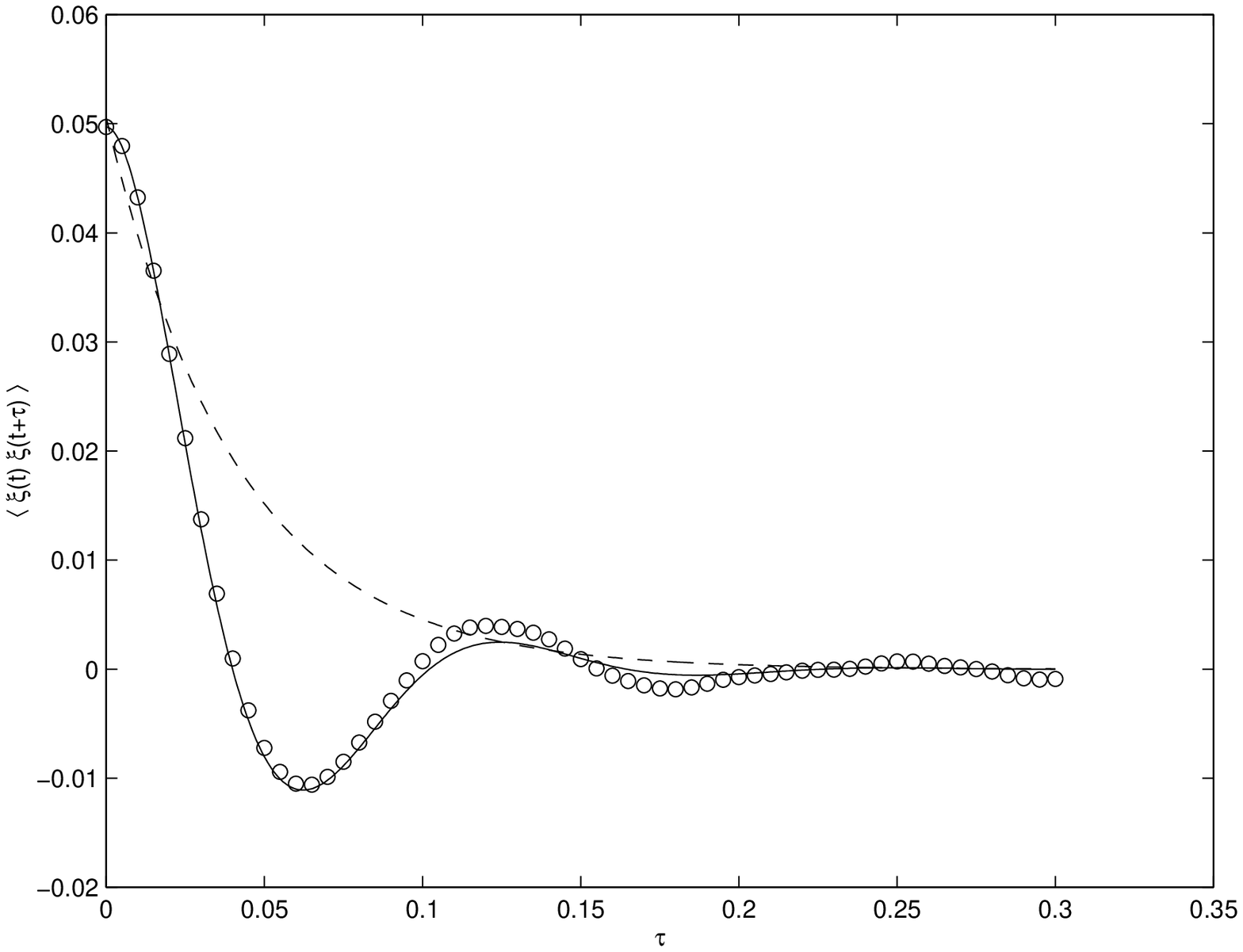}
\caption{\label{Correlation2} PDF of the torque (top) and temporal 
correlation (bottom). The points are from the
experiment, the solid line corresponds to the model OWN and the 
dashed line to the EWN model (concerning the PDF, both model give the 
same Gaussian distribution).}
\end{center}
\end{figure}

The constants appearing in the models OWN and EWN can be found by fit
on the PDF of torque measurements and correlation function. For the
model OWN, we find;
\EQA
\omega_0 &=& 55.9 \; rad.s^{-1}, \nonumber \\
\gamma &=& 24.1 \; s^{-1} \nonumber \\
D_0 &=&2 \gamma \omega_0^2 \langle \xi^2 \rangle = 7.49 \, 10^3 \; 
kg^2.m^4.s^{-7}, \\ \nonumber
c &=& 7.42 \,10^{-4} \; kg.m^2
\label{consOWN}
\ENA
while for the model EWN, we find:
\EQA
\tau &=& \frac{1}{\gamma} = 0.042  \; s \\ \nonumber
D &=& \tau  \langle \xi^2 \rangle = 2.1 \, 10^{-3} \; kg^2.m^4.s^{-3} \\ \nonumber
c &=& 7.42 \, 10^{-4} \; kg.m^2
\ENA
Our calibration can be checked by comparison between the model and
the data is provided in Fig. \ref{Correlation2}: top for the PDF and 
bottom for the
correlation function. One sees that the PDF are well reproduced with
our choice of parameter. For the correlation functions, one sees that
the model OWN captures well the first oscillation, but decreases a
little bit too slowly. 

\section{Predictions in $\Gamma$-mode}
\subsection{Numerical study of the model}
\begin{figure}[ht]
\begin{center}
\includegraphics[scale=0.42,clip]{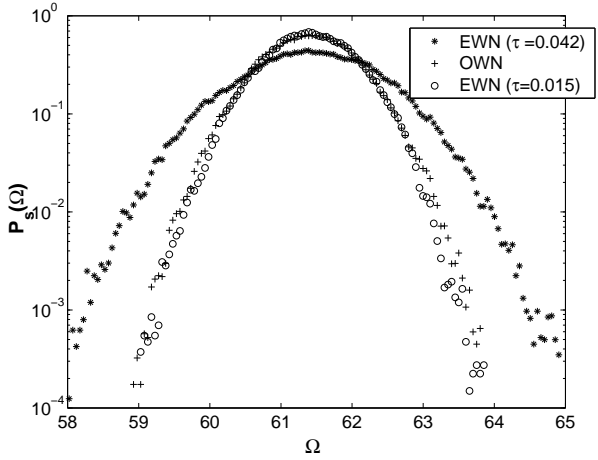}
\caption{\label{Simullog} PDF of the angular velocity computed in 
$\Gamma$-mode ($\Gamma_m =2.8\;kg.m^2.s^{-2}$) numerically for the 
two models. The parameters values are that determined in the 
calibration and $I=0.022\;kg.m^2$. We also show the numerical integration of the EWN model with values for $\tau$ et $D$ corresponding to the overdamped approximation.}
\end{center}
\end{figure}

The calibration of the model enables the determination of the
probability distribution for angular velocity in the $\Gamma$-mode.
In this case, $\Gamma_m=cte$ and the angular velocity becomes a
stochastic variable, solution of
the equation:
\EQ
\label{Gammamode}
I\frac{d\Omega}{dt}=\Gamma_m -c\vert \Omega \vert \Omega+\xi(t)
\EN
where $\xi(t)$ is given by equation (\ref{riri}) or (\ref{equaxi}).
Solutions of this coupled system of equation can be found using
classical stochastic numerical methods \cite{Kloeden92}. The only trick arises for the
OWN model whose dynamics is second order in time. In this case, we
used the following numerical scheme:
\EQA
\xi(t+\Delta t) &=& \xi(t) + y(t) \Delta t \\ \nonumber
y(t+\Delta t) &=& y(t) - \omega_0^2 \xi(t) \Delta t - 2 \gamma y(t)
\Delta t + \Gamma(t) \\ \nonumber
\Omega(t+\Delta t) &=& \Omega(t) + (\Gamma_m - c \vert \Omega(t)
\vert \Omega(t)) \frac{\Delta t}{I} + \xi(t) \frac{\Delta t}{I}
\ENA
In Fig. \ref{Simullog}, we show an example of the resulting probability
distribution function computed numerically using the two models, and
with the constants calibrated on the data. In both cases, one obtain
a PDF with a mean value $<\Omega>=61.4 \; rad.s^{-1}$, and some 
deviations from a
Gaussian character. However, one sees that the two models are
characterized by quite different variances. In a sense, this is quite
surprising because the two models can be shown to exhibit interesting
similarities under a simple approximation.

\subsection{Overdamped approximation}

Indeed, in the overdamped regime, one can neglect inertia in
(\ref{riri}) so that the noise in the OWN model obeys:
\EQA
\frac{d\xi}{dt} &=& -\frac{1}{\tau}\xi +\frac{\eta(t)}{\tau}\, ,  \\ \nonumber
\tau &=& \frac{2 \gamma}{\omega_0^2} \, , \\ \nonumber
\langle \eta(t) \eta(t') \rangle &=& \frac{2 D_0}{\omega_0^4}
\delta(t-t')= 2 D \delta(t-t')
\ENA
One therefore recovers the equation for the EWN with slightly
different parameters:
\EQA
\tau &=& \frac{2 \gamma}{\omega_0^2} = 0.015 \; s\, , \\ \nonumber
D &=& \frac{D_0}{\omega_0^4} = 7.67 \, 10^{-4} \; kg^2.m^4.s^{-3} \\ \nonumber
c &=& 7.42 \,10^{-4} \; kg.m^2
\ENA
The comparison between the PDF in the overdamped approximation (cf 
next section) and the PDF from the numerical simulation shows a good 
agreement. The present result therefore suggests that the large
difference between the two models only comes from the different
physical parameters. To understand the origin of the difference, we
undertake an analytical investigation, using the EWN model.

\subsection{Analytical study}
\subsubsection{Power in $\Omega$-mode}

In this regime, the power delivered by the propellers and the power
injected in the flow are equal. In the sequel, we shall call this
power $P_\Omega=P_1=P_2=c\Omega^3+\xi\Omega$. We can
immediately derive the PDF of $P_\Omega$ from $\xi$. It is
a gaussian random variable with mean $c\Omega^3$and variance:
\EQ
\delta P_\Omega^2 = \Omega^2 \langle \xi^2 \rangle = \frac{D
\Omega^2 }{\tau}
\label{deltaPomega}
\EN

\subsubsection{Power in Gammamode}

To find the power in the gamma-mode, one must solve analytically the 
equation (\ref{Gammamode}) with (\ref{equaxi}).
A technical difficulty arises because $\xi$ is not a
$\delta$-correlated process . However, under the unified colored
noise approximation \cite{Jung87,Ke99}, one can compute the
stationary PDF of $\Omega$ and get the following (cf appendix
\ref{UCNA}):
\EQA
\label{Pstat}
P_s(\Omega) &=& N(I+2c\tau\vert \Omega \vert) \times \\ \nonumber
\exp\frac{1}{D}[I\Omega(\Gamma_m
&-&c\Omega^2\theta(\Omega)/3)+ c\tau
\Omega^2(\Gamma_m\theta(\Omega)-c\Omega^2/2)]
\ENA
where $\theta$ is the sign function.
The moment of this distribution cannot be computed analytically in
general. The simplest approximation which allows analytical
calculations is when the intensity of the noise is small, a
regime that will be considered in the next section.

\paragraph{The  small noise limit}
In this section we rewrite the probability density function for
$\Omega$ in a dimensionless form: with $\chi=\sqrt{\frac{c}{\Gamma_m}}\Omega$,
$R^2=\frac{2D}{\tau\Gamma_m^2}$
and $S=\frac{2I}{\sqrt{c\Gamma_m}\tau}$, one has for the stationary
probability density of $\chi$:
\EQA
\label{PDFadim}
P_s(\chi)=N(\vert\chi\vert+\frac{S}{4})
\exp(-\frac{1}{R^2}[(\chi^2-\theta(\chi))^2 \\ \nonumber
-S\chi+\frac{S}{3}\theta(\chi)\chi^3])
\ENA
where N stands for the normalization.

\begin{figure}[ht]
\begin{center}
\includegraphics[scale=0.5]{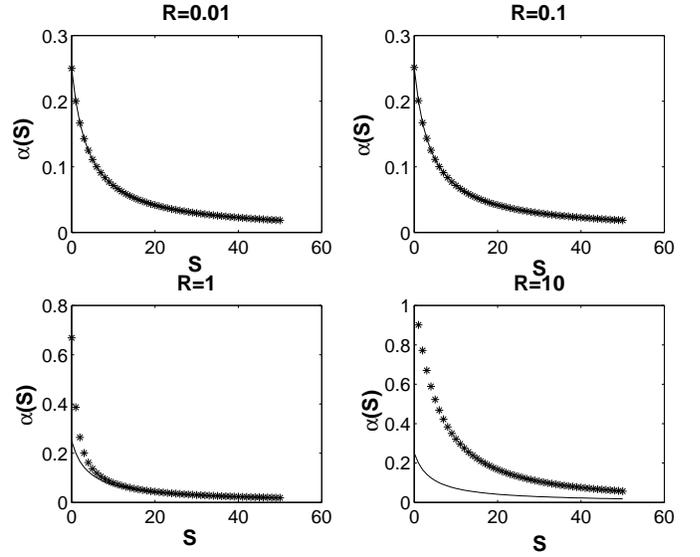}
\caption{\label{Ivariation} Evolution of the parameter
$\alpha=\frac{\delta P_{\Gamma}^2}{\delta P_{\Omega}^2}$ with S (i.e. the
inertia) for different values of $R$ (intensity of the noise. The
points correspond to the numerical computation and
the solid line to the approximate expression derived assuming the
intensity of the noise was small.}
\end{center}
\end{figure}

In the appendix \ref{moments}, we used Laplace's method with $R \ll
1$ to compute the n-th order moment of the distribution (\ref{PDFadim}):
\EQ
<\chi^n>=1-\frac{R^2}{4(4+S)}n(2-n)+O(R^4)
\EN
   From this expression, we are able to compute the standard deviation
of the processus and recover a relation enlightened
by \cite{Titon03a}. In the limit of inertia (or equivalently S) going
to zero, we have:
\EQA
\label{facteur2}
\delta P_{\Gamma}^2&\equiv&\Gamma_m^2 [<\Omega^2>-<\Omega>^2]\\ \nonumber
&=&\Gamma_m^2 <\Omega>^2[\frac{<\chi^2>}{<\chi>^2}-1]\\ \nonumber
&=&\Gamma_m^2
\frac{R^2}{8}<\Omega>^2=\frac{D}{4\tau}<\Omega>^2=\frac{1}{4}\delta
P_{\Omega}^2
\ENA
This relation shows that in the limit where the inertia of the disk
is going to zero, the fluctuations of power delivered
by the motor are twice smaller in one of the mode of forcing, namely
the $\Gamma$-mode as compared to the $\Omega$-mode
(with the same mean angular rotation rate). However, this relation
has been derived under the assumption of vanishing noise. We now have
to check if this relation holds when the noise becomes stronger and
stronger. On
figure \ref{Ivariation}, we plotted the quantity
$\alpha=\frac{\delta P_{\Gamma}^2}{\delta P_{\Omega}^2} = \frac{2<\chi^2>-<\chi>^2}{R^2<\chi>^2}$ numerically computed
from
the expression (\ref{PDFadim}) versus the adimensionalized inertia, S
and simultaneously, the expression derived in appendix
\ref{moments}, in the limit $R \ll 1$, $\alpha=\frac{1}{4+S}$. We
see roughly that for $R < 1$, the preceding relation is
in good agreement with the numerical calculations, whereas for $R >
1$, the two quantities diverge one from another.
Furthermore, in this last situation, $\alpha$ is not close to
$\frac{1}{4}$ as $S$ is going to zero. The relation
(\ref{facteur2}) is consequently only valid under the double
assumption: $R \ll 1$ and $I \rightarrow 0$.\\

In summary, we found in this section a relation linking the
fluctuations of the injected power in two forcing regimes: one with
constant velocity and one with constant
torque. This relation is valid for vanishing
inertia and states (eq. (\ref{facteur2}) )  that the power
fluctuation at constant velocity are twice
larger as power fluctuations at constant torque. This relation has
been empirically
discovered by Titon and Cadot \cite{Titon03a} using experimental
data, without discussion of its range of validity. With our model, we
predict that this range is restricted to weak noise and weak inertia,
as can be seen in figure \ref{Ivariation}.

\paragraph{Application}
Our analytical computation can be used to explain the difference
between the two models. For this, we can compute the values of the
parameter $R$ and $S$ for the two models, using the calibrated
constants. In the model OWN, we find $R=0.11$ and $S=62.6$. This indeed
corresponds to the weak noise limit, and from (\ref{casert}) we find
$\delta \Omega^2 = [\langle \chi^2 \rangle - \langle \chi \rangle^2] \Gamma_m / c = 0.34$. In the model EWN, we find $R=0.11$ and $S=23.3$. We
therefore recover the same value for the noise intensity, but a quite
different value for the adimensionalized inertia $S$. Using formula
(\ref{casert}), this results in $\delta \Omega^2 = 0.84$, about twice bigger 
as the value for the OWN
model ! This therefore explains the difference between the two models.
It is therefore now interesting to compare this PDF with experimental
data, to see which model is closer to the real distribution.

\begin{figure}[ht]
\begin{center}
\includegraphics[scale=0.45,clip]{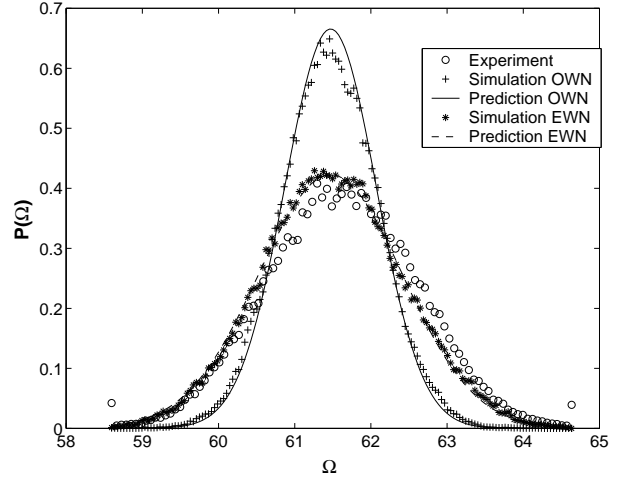}
\caption{\label{PDFOmega} Probability distribution for the angular
velocity when the device is forced at constant torque
($\Gamma_m=2.8\;kg.m^2.s^{-2}$) for the experiment, a simulation of
equations (\ref{Gammamode}) and (\ref{equaxi}), and for the
theoretical predictions (\ref{Pstat}) in the overdamped case.}
\end{center}
\end{figure}

\begin{figure}[ht]
\begin{center}
\includegraphics[scale=0.45,clip]{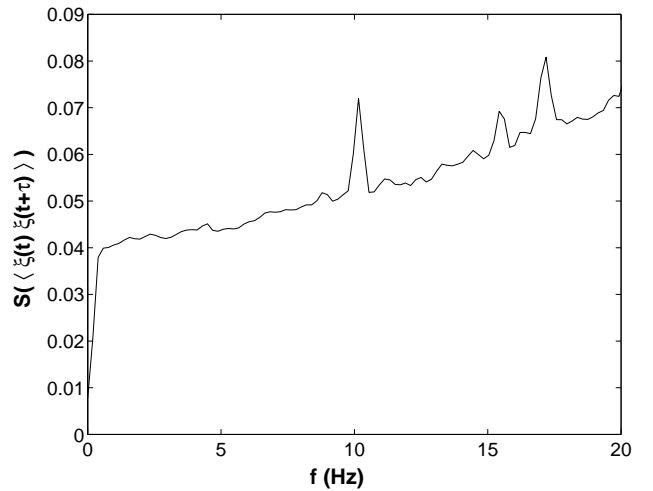}
\caption{\label{Spectre2bis} Spectrum of the correlation function of $\xi$ calculated in the $\Gamma$-mode.}
\end{center}
\end{figure}

\paragraph{Comparison with experimental data}
On figure \ref{PDFOmega}, we show the experimental data of the VKE
experiment forced with a constant torque
($\Gamma_m=2.8\;Kg.m^2.s^{-2}$) and in such a way that the mean
angular velocity is roughly $60\;rad.s^{-1}$ (the exact value is 
$61.6\;rad.s^{-1}$). This is compared with the theoretical prediction 
for both the
OWN and the EWN model.
As one can see, the experimental curve agrees very well with the EWN
model, but not with the OWN model.
A OWN model can reproduce the data only provided a change of the parameters of the fit, e.g. a oscillation's frequency smaller by a factor of $\sqrt{3}$, or no oscillation at all. Indeed the spectrum of the correlation function of $\xi$ in the $\Gamma$-regime shows no preferred frequency (figure \ref{Spectre2bis}).

In previous experiments of Von-Karman swirling flow, it has been
observed that the skewness in $\Omega$ and $\Gamma$ mode is of
different sign. It is not possible to reproduce such a feature in our
model since, by assumption, the skewness is zero in the $\Omega$ mode
(gaussian noise). However, it is possible to investigate the skewness
in $\Gamma$ mode. To proceed, one needs to carry the Laplace method
ut to terms in $R^4$ (one can check that the skewness vanishes up to
order 2) and after straightforward calculations, it appears that the
skewness is:
\EQ
\frac{<(x-<x>)^3>}{(<x^2>-<x>^2)^{3/2}}=-\frac{R(12+S)}{\sqrt{2(4+S)^3}}
\EN
This quantity is always negative, a result consistent with previous
experimental observations (see for example \cite{Pinton99}). In our
experimental result, the skewness is measured to be -0.026. With the
value of the parameters, our analytical prediction gives a value for
the skewness of -0.019 for the EWN model, and -0.011 for the OWN model.
Again, there is much better agreement between the experimental 
results and the EWN model, than with the OWN model in this regime. 
There are two possible explanation: 
\begin{enumerate}
\item the oscillation detected is a pure experimental artefact, so that OWN has no physical origin
\item the frequency of oscillation is not an universal parameter and varies accordingly to the forcing regime.
\end{enumerate}
Datas seem to select the second explanation. It is nevertheless interesting to understand
better the physical difference between the two models, trying to
pin-point its origin from the Navier-Stokes equations.

\section{What can be said from a quasi-linear model of turbulence? }
\label{modele}
\subsection{Basic equations}

In the previous sections, we have derived a Langevin model for the 
turbulent torque
$\Gamma_f$. In fact, this torque can be simply related to a component
of the Reynolds stress through the angular momentum conservation, see
e.g. Mari\'e and Daviaud~\cite{Marie03b}:
\EQ
\Gamma_f=\int_{\Sigma_p} \rho u_z u_\phi r dS
\label{torque2}
\EN
Here $\Sigma_p$ is the cross-section of the cylinder which closes the
portion
that is swept by the blades of the stirrer, and $u_\phi$
and $u_z$ denote the azimuthal and vertical fluid velocity component.
If we now separate the velocity into its mean $<u>$ and fluctuating
$u'$ contribution, we get:
\EQA
\Gamma_f=\int_{\Sigma_p} \rho (<u_z> <u_\phi>+ <u_z> u'_\phi \\ \nonumber
+u'_z <u_\phi>+ u'_z u'_\phi )r dS.
\label{torque3}
\ENA
The average of this expression gives the mean torque as:
\EQ
<\Gamma_f>=\int_{\Sigma_p} \rho (<u_z> <u_\phi>+ <u'_z u'_\phi> )r dS.
\label{torque4}
\EN
This is a classical expression for the torque transport. It is easy
to express it on pure dimensional ground as $<\Gamma_f>=-c\vert
\Omega\vert \Omega$, where $c$ is a drag coefficient. Our main
interest is the fluctuating part of $\Gamma_f$, which will provide
the noise contribution. It is:
\EQ
\Gamma'_f=\int_{\Sigma_p} \rho  (<u_z> u'_\phi +u'_z <u_\phi>+ u'_z
u'_\phi-<u'_z u'_\phi> )r dS.
\label{torque5}
\EN
At this point, it is natural to assume that the difference $u'_z
u'_\phi-<u'_z u'_\phi>$ (the fluctuating part of the Reynolds stress)
is small compared to the other two terms, which are both proportional
to the average of a mean quantity. Also, since the fluctuating part
varies over time scale much smaller than the mean part, it is easy to
see that we must have:
\EQ
D_t \Gamma'_f\approx \int_{\Sigma_p} \rho  (<u_z> D_t u'_\phi +D_t u'_z
<u_\phi>)r dS.
\label{torque6}
\EN
A model for the fluctuating torque variation will then be found
provided one finds a model for the fluctuating velocity variations.

\subsection{The quasi-linear approximation}

To obtain the
dynamical behavior of fluctuating velocities, we use the turbulent
model of Laval,
Dubrulle and Mc Williams \cite{Laval03} in which the velocity is given
as a solution of a linear stochastic equation of Langevin type, valid
for localized wave-packets, which may be summarized as:
\EQ
D_t \hat u'_i
=
-\nu^t {\bf k}^2 \hat u'_i+B_{ij}\hat u'_j+\eta_i,
\label{ssEqmod}
\EN
where $B_{ij}$ is a linear operator depending only on the average
velocity, $\nu_t$ is a turbulent viscosity, and
\EQ
{\hat u'}({\bf x},{\bf k},t)=\int g(\vert
{\bf x-x'}\vert)e^{i{\bf k\cdot (x-x')}}
{\bf u'}({\bf x'},t)d{\bf x'},
\label{gabordef}
\EN
$g$ being a function which decreases rapidly at infinity. Eq.
(\ref{gabordef}) is a Gabor transform, defining a localized
wave-packet at position $x$ with local wavenumber $k$. The advantage
of considering Gabor mode is that it allows simple treatment of
dissipation and pressure terms \cite{Laval01}. Note that by
construction, $u_i=g(0)\int dk \hat u_i(k)$.
Here, $\eta_i$ is a  noise, representing  the input of energy via the energy
cascade.
The major approximation of the model is to lump the non-linear terms
describing local interactions
into a  turbulent
viscosity $\nu^t$.\

\subsection{Reynolds stresses in the quasi-linear approximation}
Using the slow variation of $<u>$, we may write the Gabor transform
of $Q_{ij}=<u_i>u'_j$ as $<u_i>\hat u'_j$. It is then easy to see
that the Gabor transform of $Q$ satisfies, in matrix notation:
\EQ
D_t \hat Q
=
-\nu^t {\bf k}^2 \hat Q+\hat Q B^++H,
\label{ssEqmodmod}
\EN
where $H$ is a noisy matrix $H_{ij}=<u_i>\eta_j$ and the symbol $+$ 
means transposed. Decomposing finally $B$ into its symmetric part $S$ 
and
anti-symmetric part $A$, one obtains finally:
\EQA
D_t \hat P
&=&
-\nu^t {\bf k}^2 \hat P+\hat Q S+S\hat Q^++A \hat Q^+-\hat Q
A+H+H^+,\nonumber\\
D_t \hat M
&=&
-\nu^t {\bf k}^2 \hat M+\hat Q S-S\hat Q^++A \hat Q^++\hat Q A+H-H^+,
\label{ssEqmodtrois}
\ENA
where $P=Q+Q^+$ and $M=Q-Q^+$. One can get physical insights of this
system by considering the special case when $Q$ commutes with $S$ and
$A$. In that case, one gets:
\EQA
D_t \hat P
&=&
-\nu^t {\bf k}^2 \hat P+S P- A P+H+H^+,\nonumber\\
D_t \hat M
&=&
-\nu^t {\bf k}^2 \hat M+S M+A M+H-H^+.
\label{ssEqmodquatre}
\ENA
It is then easy to see that $P$ and $M$ will behave like a damped
oscillator with noise, with damping given by eigenvalues of $S-\nu^t
{\bf k}^2I$ and oscillation given by square root of eigenvalues of
$A^2$. \

\subsection{Application to torque in von Karman}
Since $\Gamma_f=\int P_{\phi z}$, we can now use the result on the
Reynolds stresses to understand the physical origin, if any, of the
various terms appearing in our model (\ref{riri}) and ({\ref{equaxi}).
We see that the friction term arises from a combination of turbulent
viscosity and symmetrical part of $B$, i.e. the mean flow stretching.
The noise term arises from the energy cascade from large to small
scale, while the possible oscillating behavior arises from the
anti-symmetrical part of $B$, i.e. is linked with the mean flow
vorticity. For example, if one approximate the von Karman flow by a
pure rotating shear flow $<u>=\sigma rz e_\phi$. Its symmetrical
tensor $S$ has only two non-zero component, $S_{\phi z}=S_{z
\phi}=0.5 r\sigma$ while the four non-zero components of $A$ are
$A_{r\phi}=-A_{\phi r}=-r\sigma$ and $A_{z\phi}=-A_{\phi z}=-0.5
r\sigma$. In that case, $A^2_{z\phi}=0$ and one can reasonably expect
that the same component of $P$ has no oscillatory behavior. This
would favor the model EWN (simple damped noise). In realistic von
Karman flow, however, a poloidal velocity component is present, due
to Ekman pumping. Imagine then that this poloidal field is able to 
couple linearly $u'_z$ and $u'_\phi$ through a term like:
\EQ
D_t \hat u'_z=\alpha \hat u'_\phi.
\label{alphaeffect}
\EN
Since $u'_\phi$ is coupled to $u'_z$ via the differential rotation:
\EQ
D_t \hat u'_\phi=\frac{d<u_\phi>}{dz}\hat u'_z,
\label{alphaeffectbis}
\EN
this induces a possible oscillatory behavior for $\hat u'_z$ and 
$\hat u'_\phi$, hence for $\Gamma_f'$.

\section{Discussion}
\label{Conclusion}
In this paper, we studied the injected power in a turbulent device, namely the von Karman swirling flow, by mean of a stochastic model of the turbulent torque. Within this frame, we obtained a few salient results, which
were tested and validated on experimental data. Assuming a Gaussian shape with exponential time-correlation of the
turbulent torque (EWN model), we recovered the link between variances of power fluctuations in two different forcing
regimes (at constant angular velocity and constant applied torque). Moreover, the model was shown to allow for parameter-free prediction of the shape of the PDF of power fluctuations in the case
with forcing at constant torque. Further experimental tests of the
model are warranted, regarding for example the statistics of the
power injected by the turbulence, or the dependence of the model
parameter with global quantities. This is left for future work.\\

\begin{acknowledgement}
We thank O. Cadot and C. Titon for having motivated the present
"two-mode" study by experimental considerations and their numerous
remarks. We wish also to thank F. Daviaud for his valuable comments
and encouragements.
\end{acknowledgement}

\appendix
{\bf APPENDIX}

\section{Derivation of the Pdf for the angular velocity}
\label{UCNA}
In this section, we show how to derive the probability distribution
function (Pdf) for $\Omega$ verifying the following
Langevin equation:
\begin{center}
\EQA
\frac{d\Omega}{dt} &=& \frac{\Gamma_m-c\vert \Omega \vert
\Omega}{I}+\frac{\xi}{I} \nonumber\\
\frac{d\xi}{dt} &=& -\frac{1}{\tau}\xi +\frac{\eta(t)}{\tau},
\ENA
\end{center}
where $\eta$ is a gaussian $\delta$-correlated noise.\

This is a non Markovian process which means that no Fokker-Planck
equation for the associated
distribution can be derived. To overcome this difficulty we used the
unified colored noise
approximation \cite{Jung87,Ke99} which permits to rewrite the
stochastic system in the following
way:
\EQ
\frac{d\Omega}{dt} = \frac{\Gamma_m-c\vert \Omega \vert
\Omega}{\varepsilon(\Omega)}+\frac{\eta(t)}{\varepsilon(\Omega)}
\EN
with $\varepsilon(\Omega) = I+2c\tau\vert \Omega \vert$. This last
equation is now Markovian and we can
instantaneously derive a Fokker-Planck equation:
\EQA
\frac{\partial P(\Omega,t)}{\partial t} = -\frac{\partial}{\partial
\Omega}(\frac{\Gamma_m-c\vert \Omega \vert \Omega}{\varepsilon(\Omega)}
P(\Omega,t))\\ \nonumber
+D\frac{\partial}{\partial \Omega}
\frac{1}{\varepsilon(\Omega)}\frac{\partial}{\partial \Omega}
\frac{1}{\varepsilon(\Omega)}
P(\Omega,t)
\ENA
The stationary distribution follows immediately by integration
(equation (\ref{Pstat})).

\section{Moments of the distribution}
\label{moments}
We want to compute the moment:
     $$
<\chi^n>=N\int_{-\infty}^{+\infty}f_n(t)\exp[-\frac{1}{R^2}\Phi(t)]dt
$$
with:
\EQA
f_n(t)=t^n(\vert t \vert+\frac{S}{4})\\
\Phi(t)=(t^2-\theta(t))^2-St+\frac{S}{3}\theta(t)t^3
\ENA

Using Laplace's method up to the second order (cf \cite{Bender}), we
have in the limit $R \ll 1$:
\EQA
\nonumber
<\chi^n>=N\sqrt{\frac{2\pi R^2}{\Phi''(t_0)}}
e^{-\frac{\Phi(t_0)}{R^2}}[f_n(t_0)-R^2(-\frac{f_n''(t_0)}{2\Phi''(t_0
)} \\ \nonumber +\frac{f_n'(t_0)\Phi'''(t_0)}{2[\Phi''(t_0)]^2}
+\frac{f_n(t_0)\Phi''''(t_0)}{8[\Phi''(t_0)]^2}-\frac{5f_n(t_0)[\Phi''
(t_0)]^2}{24[\Phi'''(t_0)]^3})] \nonumber
\ENA
where $t_0$ is the minimum of $\Phi$ on $]-\infty +\infty[$. For
every S, one can show that $t_0=1$ and using the fact
that $<\chi^0>=1$, the following expression is computed:
\EQ
\label{momentn}
<\chi^n>=1-\frac{R^2}{4(4+S)}n(2-n)+O(R^4)
\EN
For n=2, one get $<\chi^2>=1$, which is equivalent to
$<\Omega^2>=\Gamma_m/c$, a trivial relation because it's only the
mean part of equation (\ref{Gammamode}). More interesting, one can
compute the standard deviation of $\chi$ and look at its limit when
$S$ (or the inertia) tends to zero:
\EQ
\label{casert}
\frac{<\chi^2>-<\chi>^2}{<\chi>^2}=\frac{R^2}{2(4+S)}+O(R^4)
\longrightarrow \frac{R^2}{8}
\EN
When writing this last equation, in terms of $\Omega$, we recover
the equality (\ref{facteur2}).

\bibliography{VKStorque}
\bibliographystyle{plunsrt}

\end{document}